\documentclass[]{aa}

\usepackage{graphicx}% Include figure files
\usepackage{txfonts}
\usepackage{natbib}
\usepackage{booktabs}
\usepackage{hyperref}   % use for hypertext links, including those to external documents and URLs
\bibpunct{(}{)}{;}{a}{}{,} % to follow the A&A style
\newcommand{\kms}{\mbox{${\rm km\,s}^{-1}$}}

\newcommand{\Mjup}{\mbox{${\rm M}_\mathrm{J}$}}

\newcommand{\Rjup}{\mbox{${\rm R}_\mathrm{J}$}}

\newcommand{\target}{WASP-172~b\xspace}
\newcommand{\kpvsys}{$K_p-V_{\rm sys}$\xspace}
\newcommand{\numpm}[3]{$#1^{#2}_{#3}$}
\newcommand{\vorb}{\num{137.8(2.2)}\xspace}
\newcommand{\vsys}{\num{-20.283(0.006)}\xspace}
\usepackage[normalem]{ulem}

\usepackage[separate-uncertainty = true,multi-part-units=single,group-separator={}, group-digits = integer, group-four-digits = true,per-mode=reciprocal]{siunitx}
\DeclareSIUnit\parsec{pc}
\usepackage{chemformula}

\begin{document}

\title{Detection of atmospheric species and dynamics in the bloated hot Jupiter WASP-172~b with ESPRESSO\thanks{Based on observations made at ESO's VLT (ESO Paranal Observatory, Chile) accessible under ESO programme 109.22Z4.006 (PI Albrecht).}}

%\titlerunning{Detection of atmospheric species and dynamics in the bloated hot Jupiter WASP-172~b with ESPRESSO}
\authorrunning{Seidel \& Prinoth et al.}

\author{J.~V.~Seidel\inst{1}\thanks{ESO Fellow, correspondence: jseidel@eso.org, joint first author}
\and B.~Prinoth\inst{1,2}\thanks{joint first author} 
\and E.~Knudstrup\inst{3,4}
\and H.~J.~Hoeijmakers\inst{2}
\and J.~J.~Zanazzi\inst{5}\thanks{51 Pegasi b Fellow}
\and S.~Albrecht\inst{3}
}
\institute{European Southern Observatory, Alonso de C\'ordova 3107, Vitacura, Regi\'on Metropolitana, Chile
\and Lund Observatory, Division of Astrophysics, Department of Physics, Lund University, Box 43, 221 00 Lund, Sweden
\and Stellar Astrophysics Centre, Department of Physics and Astronomy, Aarhus University, Ny Munkegade 120, DK-8000 Aarhus C, Denmark
\and Chalmers University of Technology, Department of Space, Earth and Environment, 412 93, Gothenburg, Sweden
\and Astronomy Department and Center for Integrative Planetary Science, University of California
Berkeley, Berkeley, CA 94720, USA
}

\date{Received date/ Accepted date}

\abstract{\textit{Context.} 

The population of strongly irradiated Jupiter-sized planets has no equivalent in the Solar System. It is characterised by strongly bloated atmospheres and atmospheric large-scale heights. Recent space-based observations of SO$_2$ photochemistry demonstrated the knowledge that can be gained from detailed atmospheric studies of these unusual planets about Earth's uniqueness. \\
\textit{Aims.} Here we explore the atmosphere of WASP-172~b a similar planet in temperature and bloating to the recently studied HD~149026~b. In this work, we characterise the atmospheric composition and subsequently the atmospheric dynamics of this prime target. \\ 
\textit{Methods.} We observed a particular transit of WASP-172~b in front of its host star with ESO's ESPRESSO spectrograph and analysed the spectra obtained before during and after transit.\\
\textit{Results.} We detect the absorption of starlight by WASP-172~b's atmosphere by sodium ($5.6\sigma$),  hydrogen ($19.5\sigma$) and obtained a tentative detection of iron ($4.1\sigma$). We detect strong - yet varying  - blue shifts, relative to the planetary rest frame, of all of these absorption features. This allows for a preliminary study of the atmospheric dynamics of WASP-172~b. \\
\textit{Conclusions.} With only one transit, we were able to detect a wide variety of species, clearly tracking different atmospheric layers with possible jets. WASP-172~b is a prime follow-up target for a more in-depth characterisation both for ground and space-based observatories. If the detection of Fe is confirmed, this may suggest that radius inflation is an important determinant for the detectability of Fe in hot Jupiters, as several non-detections of Fe have been published for planets that are hotter but less inflated than \target.
}


\maketitle

%----------------------------------------------------------------------------------------
%       ARTICLE CONTENTS
%----------------------------------------------------------------------------------------
\section{Introduction}

In comparison to the rest of the exoplanet population, hot Jupiters remain rare and are only found around approximately $1\%$ of solar-type stars. Nonetheless, due to their intrinsically small semi-major axis, they are found ubiquitously via the transit method, including the WASP\footnote{Wide Angle Search for Planets} survey \citep{Pollacco2006}. They are characterised by strong irradiation from their host star and have no equivalent in the Solar System. In the sample of hot Jupiters, less than two dozen have unequivocally confirmed atmospheres. Bloated hot Jupiters, with their large-scale heights, make outstanding targets to further our knowledge of these strange objects.

\begin{figure}
 \centering
 \label{fig:massinso}
\resizebox{\columnwidth}{!}{\includegraphics[trim=0.0cm 0.0cm 2.5cm 0.0cm]{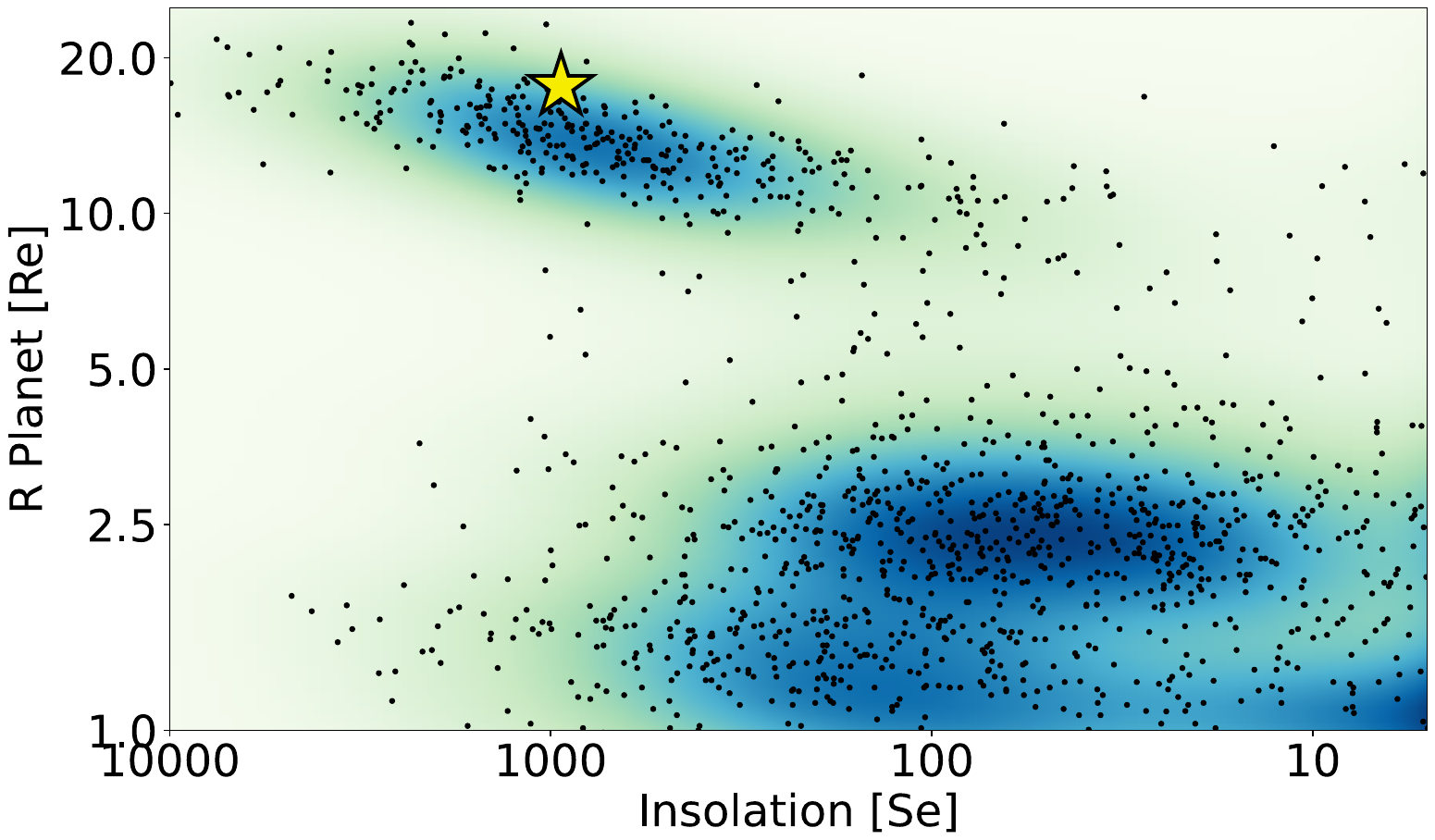}}
 \caption{The bloated hot Jupiter \target{} is shown as a yellow star in the mass vs. insolation space in Earth units. The data was retrieved from the NASA Exoplanet Archive at \url{https://exoplanetarchive.ipac.caltech.edu/}.}
\end{figure}

\noindent WASP-172~b, one of the most bloated hot Jupiters found to date, is shown in the mass-insolation space in Figure \ref{fig:massinso}\citep{Hellier2019}. WASP-172~b's host star is an F1V-type star (Vmag = $11.0$, distance $\sim$\SI{530}{\parsec}) which it orbits in \num{5.48} days \citep{Hellier2019}. It is one of the more bloated planets studied to date with a mass of \SI{0.47}{\Mjup} and a radius of \SI{1.57}{\Rjup}, resulting in a mean density of only \SI{0.16(0.05)}{\g\per\cm\cubed}, see Table\,\ref{tab:parameters_W172} for an overview of the system parameters. 

In this work, we provide the first evidence of various atmospheric detections from one ESPRESSO transit for WASP-172~b, demonstrating its potential as a promising target for additional in-depth study.

This paper is structured as follows: in Section \ref{sec:dataset}, we provide an overview of the ESPRESSO dataset and discuss various possible contamination sources and corrections, followed by the detections obtained with narrow-band transmission spectroscopy (Section \ref{sec:narrowband}) and cross-correlation analysis (Section \ref{sec:ccf}). We then discuss our findings and provide an outlook for future work in Section \ref{sec:diss}.

\begin{table}[h!]
	\caption[]{Summary of stellar and planetary parameters of the WASP-172 system adopted in this study.}
	\label{tab:parameters_W172}
	\small
	\begin{center}
		\def\arraystretch{1.25}
		\begin{tabular}{p{0.52\linewidth}p{0.33\linewidth}p{0.025\linewidth}}
			\toprule
            \multicolumn{3}{c}{WASP-172 System parameters}  \\ \midrule
            RA$_{\rm 2000}$ & 13:17:44 & [2]\\
                DEC$_{\rm 2000}$ & -47:14:15 & [2]\\ 
                Parallax $p$ [mas] & \num{1.8942(0.0222)} & [2] \\
                Magnitude [V$_{\rm mag}$] & 11.0 & [2]\\
                Systemic velocity ($v_{\rm sys}$) [\si{\km\per\second}]    & \num{-20.283(0.006)} & [1] \\
                 \midrule
                \multicolumn{3}{c}{Stellar parameters} \\ \midrule
                Star radius ($R_\ast$) [$R_{\odot}$]	& \num{1.91(0.10)} & [1]\\
			Star mass ($M_\ast$) [$M_{\odot}$] & \num{1.49(0.07)} & [1] \\
			Proj. rot. velocity ($v\sin{i}$) [\si{\km\per\second}] & \num{13.7(1.0)} & [1] \\
                Age [Gyr] & \numpm{1.2}{0.0}{-1.2} & [1] \\
                Metallicity [Fe/H] & \num{-0.1(0.08)} & [1] \\
			\midrule
			\multicolumn{3}{c}{Planetary parameters}  \\ \midrule
			Planet radius ($R_{\rm p}) $ [$R_{\rm Jup}$]  & \num{1.57(0.10)} & [1]    \\
			Planet mass ($M_{\rm p} $) [$M_{\rm Jup}$]  & \num{0.47(0.10)} & [1] \\
			Eq. temperature ($T_{\rm eq}$) [$\si{\kelvin}$] & \num{1740(60)} & [1] \\
			Density ($\rho$) [\si{\g\per\cm\cubed}] & \num{0.16(0.05)} & [1] \\
			Surface gravity ($\log g_{\rm p}$) [cgs] & \num{0.12(0.04)} & [1] \\
			\midrule
			\multicolumn{3}{c}{Orbital and transit parameters} \\ \midrule
			Transit centre time ($T_0$) [HJD (UTC)]  & \num{2457032.2617(0.0005)} & [1] \\
			Orbital semi-major axis ($a$) [au]   & \num{0.0694(0.001)} & [1] \\
			Scaled semi-major axis ($a/R_\ast$)  & \num{8.0(0.5)}& [1]\\
			Orbital inclination ($i$) [$^\circ$]   & \num{86.7(1.1)} & [1]  \\
			Projected orbital obliquity ($\lambda$) [$^\circ$] & \num{121(13)} & [3] \\
			Eclipse duration ($T_{14}$) [h] & $5.294 \pm 0.048$ & [1]\\
			Radius ratio ($R_p^2/R_\ast^2$) & \num{0.0072(0.0002)} & [1] \\
			RV semi-amplitude ($K$)  [\si{\km\per\second}] & \num{0.042(0.009)} & [1]\\
			Period ($P$) [d]   & \num{5.477433(0.000007)} & [1]  \\
			Eccentricity & \numpm{0}{+0.28}{-0.00} & [1]  \\
			\midrule
			\multicolumn{3}{c}{Derived parameters} \\
			\midrule 
			Planetary orbital velocity ($v_{\rm orb}$)  [\si{\km\per\second}]    & \num{137.8(2.2)} & \\   
			Approx. scale height ($H$) [\si{\km}]                    & \num{1631(56)} & \\
			Transit depth of $H$ ($\delta F / F$)  {\scriptsize [$\times 10^{-5}$]}  & \num{20.9(2.7)}  & \\

			\bottomrule
		\end{tabular}
	\end{center}
 \noindent{\footnotesize{References: [1] \cite{Hellier2019}, [2] \cite{Gaia2020}, [3] Knudstrup et al. (in prep)}}
 \end{table}

\section{ESPRESSO dataset and initial analysis}
\label{sec:dataset}

\begin{figure}
    \centering
    \includegraphics[width=\linewidth]{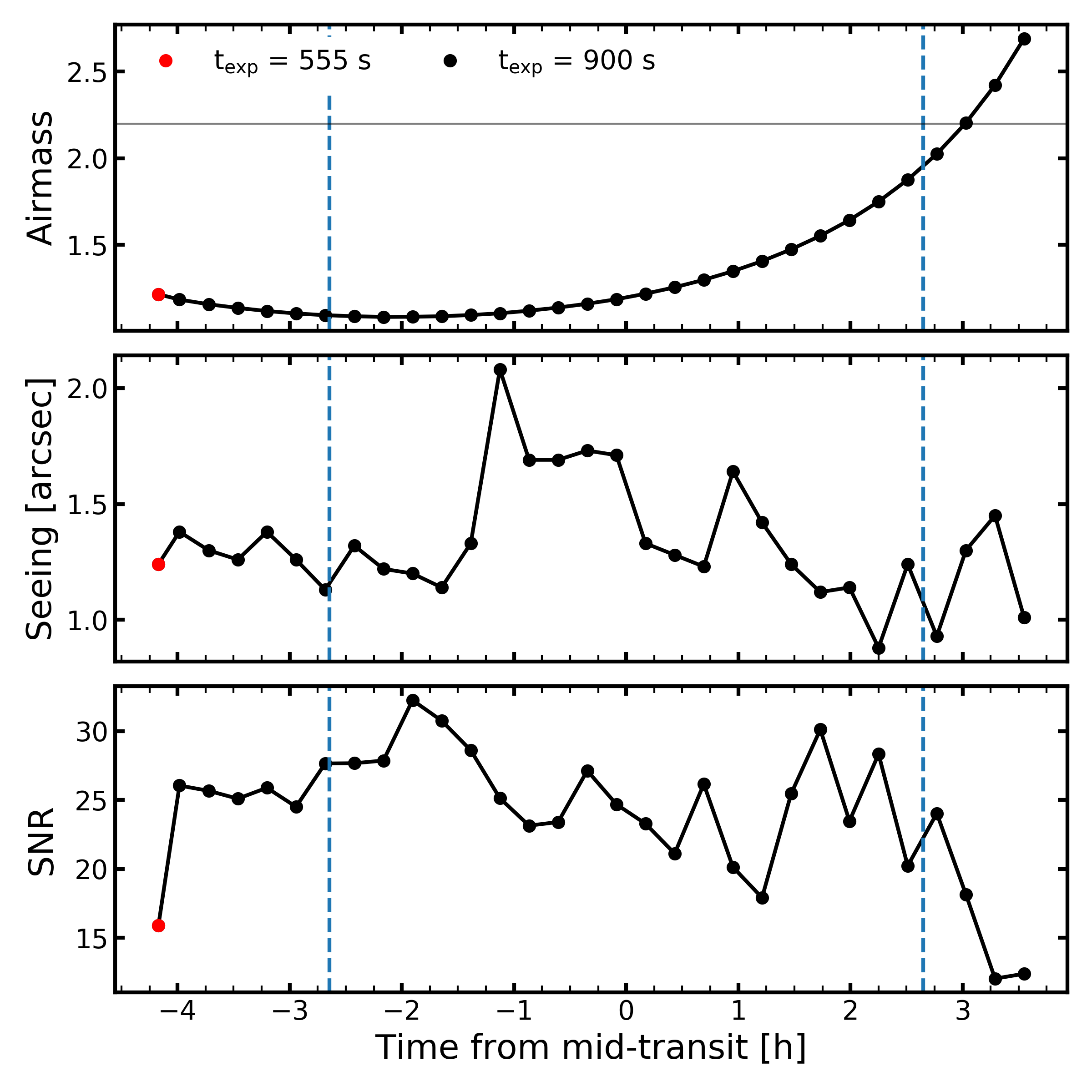}
    \caption{Log of observations. Airmass (upper), seeing (middle), and signal-to-noise ratio in order 56 (bottom) during the time series. The vertical dashed blue lines indicate the start and the end of the transit. The first exposure has a shorter exposure time of 555 s. The grey horizontal line indicates the 2.2 airmass calibration limit of the ADC based on which the three last exposures were rejected.}
    \label{fig:nightoverview}
\end{figure}

We observed the bloated hot Jupiter WASP-172~b with the ESPRESSO echelle spectrograph at ESO's VLT telescopes in Paranal Observatory, Chile \citep{Pepe2021} during the night of 2022-Jun-01 as part of ESO programme 109.22Z4.006, PI: Albrecht. Figure \ref{fig:nightoverview} shows an overview of the observed transit.

\noindent In total, 31 spectra were observed with 20 spectra taken in transit. The first spectrum was taken with a shorter exposure time of 555 s instead of \SI{900}{\second} for testing purposes and weighed by their signal-to-noise ratio accordingly. Additionally, the target was observed until the end of its visibility at Paranal with airmass 2.6, however, the ESPRESSO ADC (Atmospheric Dispersion Corrector) is only calibrated for airmass $<2.2$. Therefore, the last 3 spectra (which had an airmass above 2.2) were rejected from the analysis, leaving a total of 8 out-of-transit spectra for the narrow-band analysis, 7 before transit and one after transit. For the cross-correlation analysis, we additionally removed the shorter exposure due to a visibly higher noise level. Cosmic rays were rejected at the $5 \sigma$ level and replaced with the time-averaged mean \citep[for more information see e.g.][]{Wyttenbach2015, Seidel2019}.

Various effects have to be corrected before the planetary atmospheric signal can be extracted. In the following, we first correct for telluric lines, mainly caused by \ch{O2} around the sodium doublet and \ch{H2O} overall. We then study the impact of telluric sodium emission and correct for stellar effects.

We corrected the imprint of telluric absorption on our spectra with ESO's {\tt molecfit}, version 1.5.1. \citep{Smette2015, Kausch2015} with parameters as described in \citet{Allart2017}.
Molecfit uses measured atmospheric conditions on-site during the observations and corrects micro-telluric and stronger telluric lines. Each individual spectrum is fit independently accounting for changing seeing and airmass during the time series. The thus-created telluric line profile is then divided from each spectrum. We checked the correction by visually assessing whether the before and after master spectra show any over or under-correction of known telluric lines around the sodium doublet where various strong telluric lines are present. For the cross-correlation technique, where wide wavelength ranges are needed rather than specific lines, the wavelength regions where the correction left visible residuals were masked out manually.

\subsection{Telluric sodium emission}

For the search for sodium, assessing the contribution of the sky emission is particularly important. We monitored for telluric sodium emission or sodium laser contamination with fibre B set on sky. No strong laser contamination in the sodium D$_2$ was found. However, in both the positions of the telluric sodium D$_2$ and D$_1$ line a sodium emission peak three times the noise level was found in the fibre B spectra, most notably at the end of the observations at higher airmass. The small magnitude of the emission lines together with the airmass dependency points towards the origin of meteor excitation of atmospheric sodium \citep{Chen2020, Seidel2020c}. Our observations coincided with the end of the 2022 $\tau$-Herculids meteor shower which was visible from Paranal at roughly $20 ^\circ$ altitude just above the horizon\footnote{\url{https://spacetourismguide.com/tau-herculids-meteor-shower/}}. These most affected spectra at the highest airmass were rejected from the analysis based on the ESPRESSO ADC restrictions and do not influence our results. For the remaining contamination, the line centres of the sodium emission do not overlap with the planetary trace. When taking into account the barycentric Earth radial velocity of $\sim$\SI{-13}{\km\per\second} during the entire transit, the centre of the remaining constant contamination in the stellar rest frame lies at \SI{5889.25}{\angstrom} and \SI{5895.35}{\angstrom}, respectively. For our overall transmission spectrum, the in-transit spectra are shifted in the planetary rest frame subsequentially from \SIrange{-28}{+22}{\km\per\second}. This smears out the sodium emission contamination over a passband of roughly \SI{1}{\angstrom} around the stellar rest frame centres with no overlap with the planetary signal.

\subsection{Stellar contamination}

\begin{figure}
    \centering
    \includegraphics[width=\linewidth]{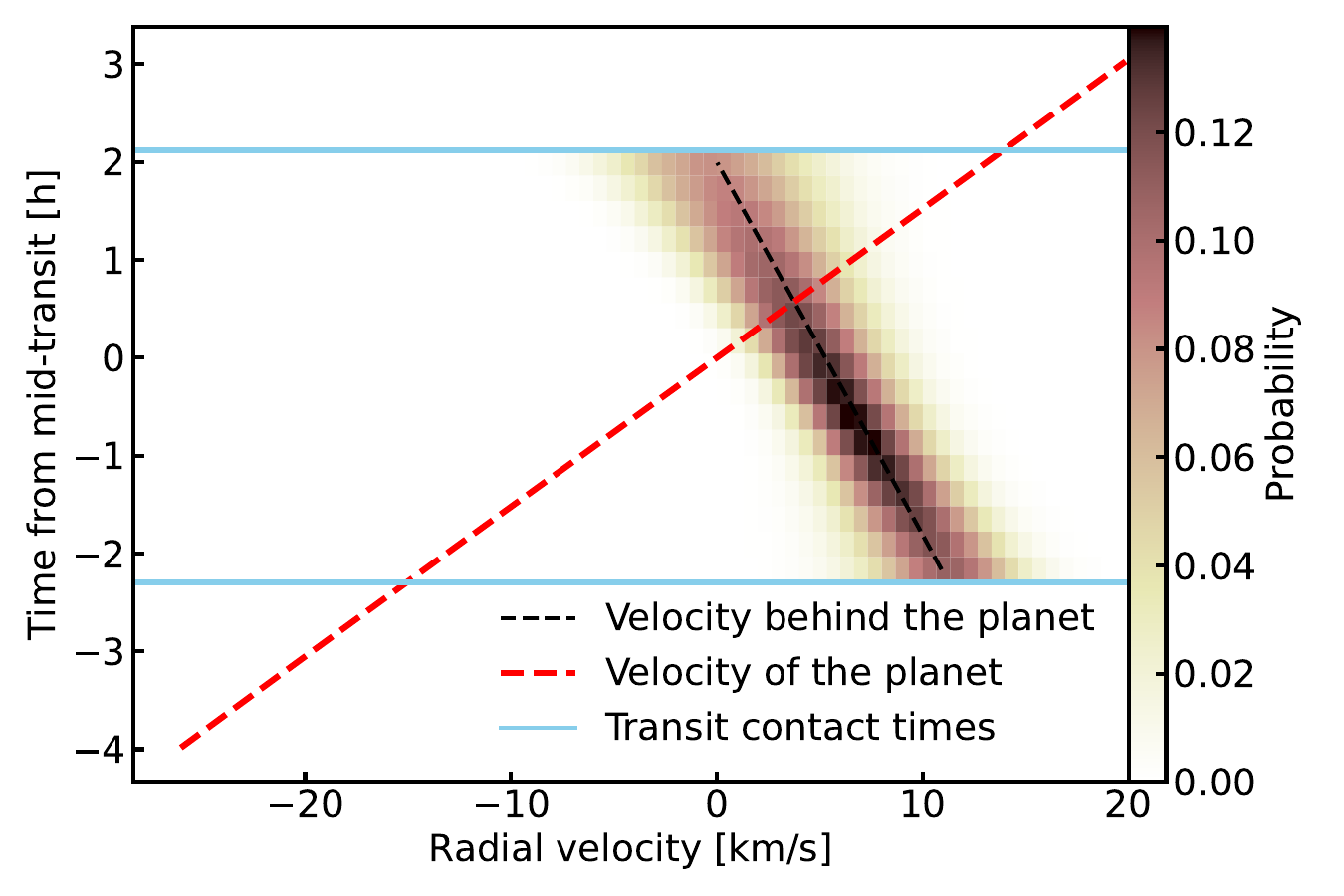}
    \caption{Radial velocity extent of the velocity component obscured by the planet during transit. The red dashed line shows the expected planetary velocity. The blue lines indicate the start and the end of the transit. The black dashed line shows the mean radial velocity extent of the velocity component obscured by the planet during transit as predicted using 100,000 samples, assuming normal distributions for the planetary and stellar parameters listed in Table\,\ref{tab:parameters_W172}.}
    \label{fig:DS_investiagtion}
\end{figure}

As the planet passes in front of the star it obscures part of the rotating stellar disk, leading to a distortion of the stellar spectral lines known as the Rossiter-McLaughlin (RM) and Centre-to-Limb variation (CLV) effects
\citep{Rossiter1924,Mclaughlin1924,Triaud2018,AlbrechtDawsonWinn2022}. To properly account for this, we need to know the projected obliquity, $\lambda$, i.e., the (projected) angle between the stellar spin axis and the orbital axis of the planet, in order to trace out the path of the planet in the local velocity field of the stellar surface. We did this by modeling the planetary shadow following the method by \citet{Albrecht2007}, specifically as outlined in \citet{Knudstrup2022}. From this we obtained a value of $\lambda=$\SI{121(13)}{\deg} (Knudstrup et al. in prep.). For narrow-band transmission spectroscopy, we then applied a numerical correction of the RM and CLV effects following \citet{Wyttenbach2020} with the local velocity field of the stellar surface to obtain the resolved planetary spectral features. For our cross-correlation analysis, as we need a correction for a much wider wavelength range, we model the radial velocity extent of the velocity component that is obscured by the planet during transit. This component is often termed Doppler shadow and marked with a black dashed line in Figure \ref{fig:DS_investiagtion}. Since the feature is not as prominent in cross-correlation space as for other systems \citep[e.g.][]{Hoeijmakers2020,Prinoth2022}, we calculate the radial velocity extent expected for the feature following \citet{Cegla2016}, see Eqs. (2) - (8) therein. Assuming normal distributions for the scaled semi-major axis $a/R_\mathrm{s}$, the orbital inclination $i$, the projected orbital obliquity $\lambda$, the projected rotational velocity $v\sin i$ and no differential rotation, we draw 100,000 samples to calculate the possible traces of the residual of the obscuration, see Figure \,\ref{fig:DS_investiagtion}. It is evident that the trace of the residual of the planetary obscuration of the stellar disc overlaps with the planetary trace for a short period of time after mid-transit. Using the expected radial velocity extent as our fitting prior, we model the feature as described in Section\,\ref{sec:ccf}.

We did not perform any checks for rotational modulations, given that no significant rotational modulations of the host star were detected that could mimic an atmospheric signal \citep{Hellier2019}.

\section{Narrow-band transmission spectroscopy}
\label{sec:narrowband}

We calculate the transmission spectrum from the ESPRESSO transit following \citet{Tabernero2020,Borsa2021,Seidel2022}, which we summarise here briefly. All spectra were weighted by their S/N and corrected for telluric lines. Then all spectra are shifted from the observer's rest frame to the stellar rest frame. In the stellar rest frame, the stellar spectral lines coincide at the same wavelengths and are removed from the magnitude-smaller planetary signal by dividing the in-transit spectra by the normalised sum of all out-of-transit spectra, the so-called master-out. The separated planetary spectra are then shifted into the planetary rest frame and combined for the final transmission spectrum. The velocities applied for the various reference frame changes are the barycentric earth radial velocity $\sim$\SI{-13}{\km\per\second}, the planet velocity ranging from \SIrange{-28}{+22}{\km\per\second}, and a negligible stellar velocity of few \si{\m\per\second}. The system velocity is \SI{-20.28}{\km\per\second} \citep[taken from][]{Hellier2019}.

 \begin{figure*}[htb]
\resizebox{\textwidth}{!}{\includegraphics[trim=0.5cm 0.0cm 0.5cm 0.0cm]{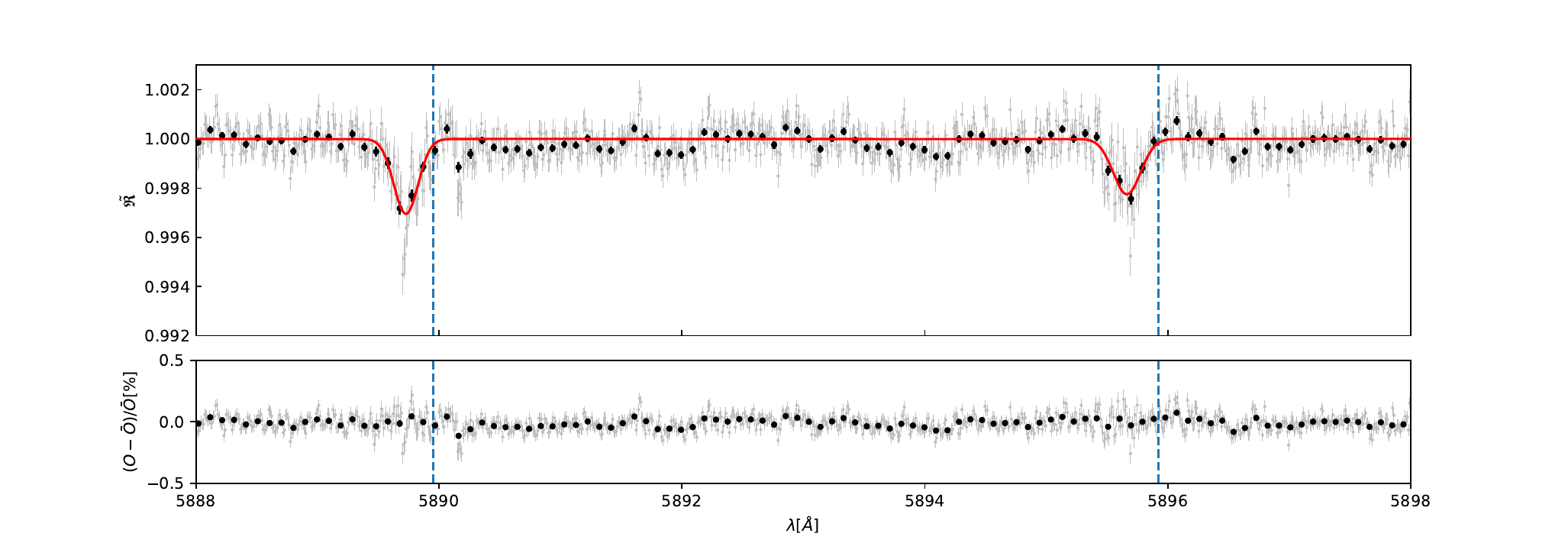}}
        \caption{ESPRESSO transmission spectrum of WASP-172~b for the sodium doublet in the planetary rest frame. Upper panel: The transmission spectrum in full resolution is shown in grey, in black the same data is shown binned by x20. The data has been corrected for tellurics, cosmics, and the RM+CLV effect. The rest frame transition wavelengths are marked with blue dashed lines. In red, a Gaussian fit to both lines simultaneously is shown. Lower panel: Residuals of the Gaussian fit. }
        \label{fig:transspectrum_Na}
\end{figure*}

\begin{figure}[htb]
\resizebox{\columnwidth}{!}{\includegraphics[trim=1.0cm 0.0cm 1.0cm 0.0cm]{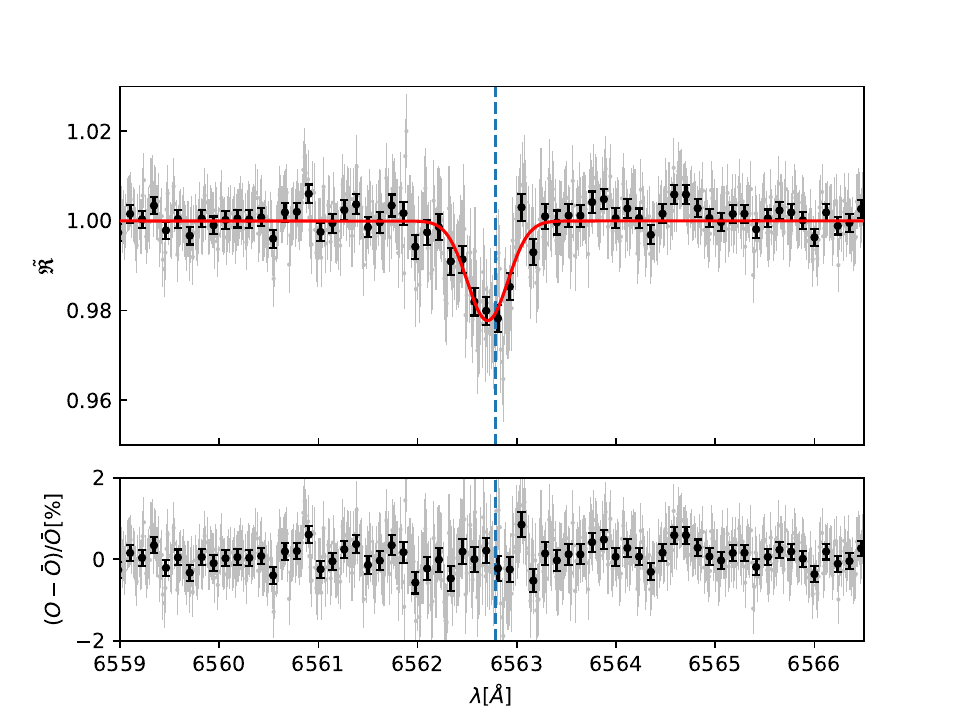}}
        \caption{ESPRESSO transmission spectrum of WASP-172~b for the wavelength range of the H-$\alpha$ line in the planetary rest frame. Upper panel: The transmission spectrum in full resolution is shown in grey, in black the same data is shown binned by x20. The data has been corrected for tellurics, cosmics, and the RM+CLV effect. The rest frame transition wavelength is marked with a blue dashed line. In red, a Gaussian fit is shown. Lower panel: Residuals of the Gaussian fit. }
        \label{fig:transspectrum_Ha}
\end{figure}

\noindent The Coud\'{e} Train optics of ESPRESSO generate interference patterns, which create sinusoidal noise in the transmission spectra at the same order of magnitude as planetary lines \citep{Allart2020,Tabernero2020}. The true origin and dependence on airmass and other factors are not yet fully understood, but efforts are currently underway to correct this behaviour on the pipeline level (private communication). In the orders of the sodium doublet and the H-$\alpha$ line, no notable wiggles were visible, however, it is likely that the much stronger low-frequency wiggles in other orders inhibited the detection of other resolved spectral lines (e.g. in the order of Li). Once the pipeline is able to mitigate this interference pattern, the analysis presented here should be expanded to include species that we could not detect. 

We show the transmission spectra of sodium and H-$\alpha$ in Figures \ref{fig:transspectrum_Na} and \ref{fig:transspectrum_Ha} combining both spectral orders independently containing the wavelength range of interest: for the sodium doublet \num{116} and \num{117}, and for H-$\alpha$ the orders \num{138} and \num{139}. The Gaussian fit shown in red for both figures is generated on the unbinned data in grey with width, depth, and position as free parameters.

\subsection{False positive assessment}
\label{sec:false_pos}

\begin{figure}[htb]
\resizebox{\columnwidth}{!}{\includegraphics[trim=0.0cm 0.0cm 0.0cm 0.0cm]{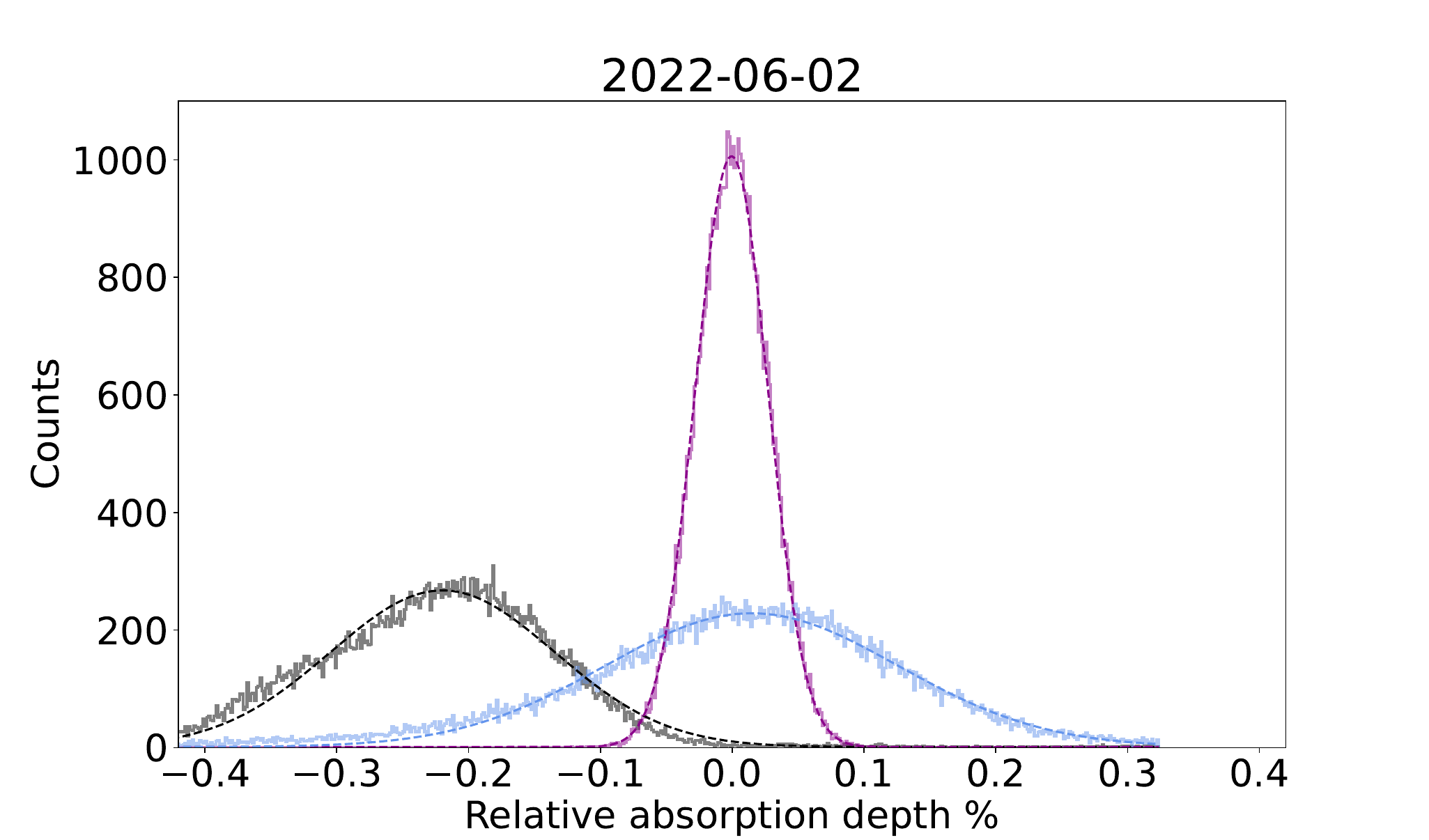}} 
        \caption{Distribution of the bootstrapping analysis for the $12$ \r{A} passband around the sodium doublet for 50,000 random selections. The `in-in' (magenta) and `out-out' (light blue) distributions are centred at zero (no planetary sodium detection) and the `in-out' distribution is shown in black indicating a planetary origin of the feature.}
        \label{fig:bootstrap}
\end{figure}

\noindent A false positive detection can arise from instrumental effects, stellar spots, or the variation of observational conditions during the night. The false positive probability can be calculated via a bootstrap analysis with an empirical Monte Carlo \citep[EMC;][]{Redfield2008}. The false positive probability is then directly taken into account when calculating the detection error.

\noindent To understand whether only the here presented assessment of in- and out-of-transit data yields a detection, in-, and out-of-transit data are considered as two independent datasets, where only the in-transit data should contain real traces of the planetary signature. Following the approach in \citet{Redfield2008}, virtual transmission spectra are created by randomly drawing new virtual in-transit and out-of-transit datasets from our real data. We distinguish between three scenarios: `in-in' (all spectra taken from the real in-transit data and randomly attributed to virtual in and out-of-transit data), `out-out' (drawn only from out-of-transit spectra), and `in-out', where virtual in-transit is drawn from real in-transit and vice versa for the out-of-transit pair. If the detection truly stems from the planetary atmosphere and is not a spurious event, only the `in-out' scenario should yield a significant result.

\noindent From this assessment, we calculate the false positive likelihood as the standard deviation of the `out-out' distribution. The `out-out' distribution is used as it does not contain any actual planetary signal and a detection in this dataset would be a true false positive. To mitigate the inherent selection bias of this method, the likelihood has to be scaled by the square root of the fraction of out-of-transit spectra to total spectra taken \citep{Redfield2008,As13}.

\noindent The bootstrapping distribution for the sodium doublet is shown in Figure \ref{fig:bootstrap}, with $50,000$ iterations. The `in-in' and `out-out' distributions are centred at $0$, ruling out spurious features as the cause for our detections. The `in-out' distribution is centred at $-0.22\%$,  highlighting that the signal origins only from the in-transit exposures and shows uniformity. This analysis also shows that the sodium detection is not influenced by the sodium emission contamination described in Section \ref{sec:dataset}. The false positive likelihood for the analysed transit is $8.5\times10^{-4}$.

\begin{figure}[htb]
\resizebox{\columnwidth}{!}{\includegraphics[trim=0.0cm 0.0cm 0.0cm 0.0cm]{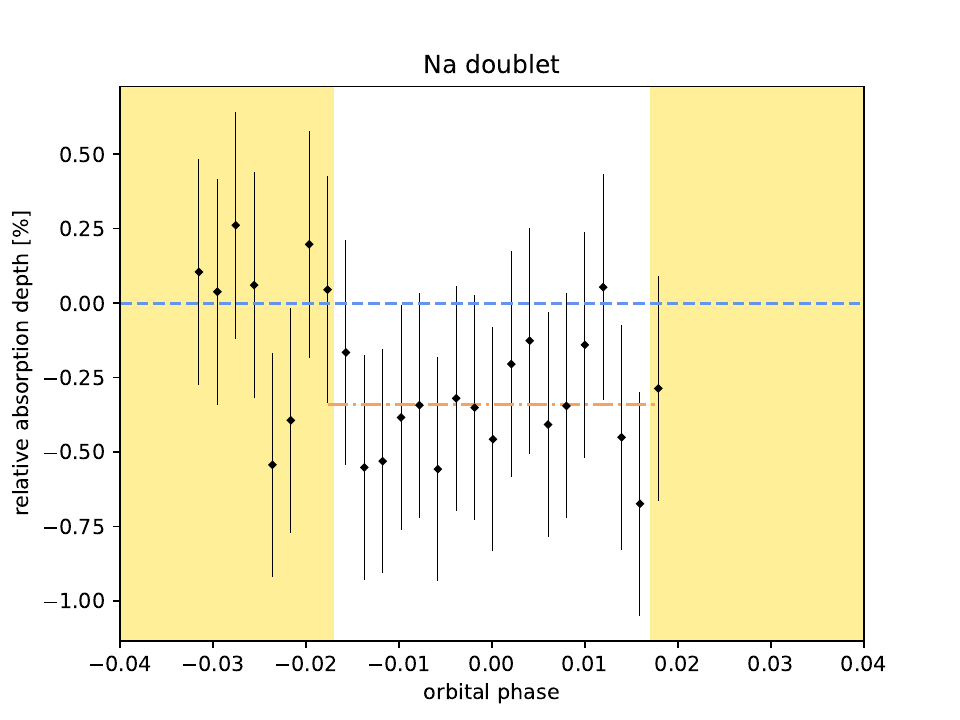}} 
        \caption{Relative absorption of the sodium doublet in the planetary restframe as a function of phase. The blue, dashed line shows the neutral $0.0$, and the brown, dashed-dotted line the in-transit mean, the shaded areas mark the out-of-transit phases.}
        \label{fig:relLightcurve}
\end{figure}

To additionally assess whether the sodium detection stems truly from the planet and not from stellar phenomena we created the relative absorption light curve for sodium (see Figure \ref{fig:relLightcurve}) showing the extra drop in flux for the in-transit spectra when integrating over the wavelength range of the sodium doublet, following \cite{Seidel2019}. We can clearly see a marked extra drop in flux during transit (highlighted by the brown, dashed-dotted mean), but it also highlights the limitations of the dataset with a short out-of-transit baseline only before the transit.

\subsection{Atmospheric detection levels}

 \begin{table}
\caption{Narrow band detection levels.}
\label{table:detections}
\centering
\begin{tabular}{ c | c c }
\hline
\hline
 & Detection [$\%$] & $\sigma$     \\
\hline
 Na D$_2$ & $0.305\pm0.067$ & $4.5$\\
 Na D$_1$ & $0.225\pm0.068$ & $3.3$ \\
combined & $0.265\pm0.048$ & $5.6$ \\
\hline
 H-$\alpha$ & $2.222\pm0.114$ & $19.5$\\
\hline
\end{tabular}
\end{table}

We searched for resolved spectral lines of Na, H-$\alpha$, Mg, Li, and K and were able to unambiguously detect the sodium doublet (shown in Figure \ref{fig:transspectrum_Na}, $5.6~\sigma$ detection for both lines combined) and the H-$\alpha$ line of the Balmer-series of hydrogen (shown in Figure \ref{fig:transspectrum_Ha}, $19.5~\sigma$ detection). Both spectral ranges are shown in the planetary rest frame and the expected line centres are indicated as dashed blue vertical lines. Gaussian fits to the spectral ranges are shown in red. The amplitude of the fit was used to estimate the absorption depth in Table \ref{table:detections} while the uncertainty is composed of the uncertainty of the Gaussian fit, the average uncertainty within the FWHM of the detected lines, and the false positive likelihood calculated in Section \ref{sec:false_pos} following \citet{Hoeijmakers2020}. 

 While the two detections are unambiguous, a study of the line shape relies on confirmation of the results presented here. There is currently no second transit of this target available as the here presented dataset is part of an exploratory programme geared towards understanding system architectures via the RM effect with only one transit per target. However, the H-$\alpha$ and sodium lines both show curiously different blue-shifted offsets which will be discussed in Section \ref{sec:dynamics} without an in-depth analysis of atmospheric dynamics.

\section{Cross-correlation analysis}
\label{sec:ccf}

We further analysed the transit time series using the cross-correlation technique \citep{Snellen2010} following the methodology in \citet{Prinoth2022}. After telluric correction as described in Section\,\ref{sec:dataset}, the individual spectra are shifted to the rest frame of the host star. The Doppler shifts include corrections for the Earth's velocity around the barycentre of the Solar System and the radial velocity of the host star caused by the orbiting planet. The velocity corrections yield a stellar spectrum with a constant velocity shift that is consistent with the systemic velocity of \SI{-20.283(0.006)}{\km\per\second}.
To account for outliers, we follow \citet{Hoeijmakers2020} and apply an order-by-order sigma clipping algorithm that computes a running median absolute deviation over sub-bands of the time series spanning 40 pixels in width. Pixels with deviations larger than $5\sigma$-outliers were rejected and interpolated. Additionally, we flagged spectral columns where the telluric correction left systematic noise, mainly caused by deep telluric lines. The rejection of outliers and manual masking of spectral pixels affected 7.81\% of the pixels in the time series.
Using cross-correlation templates at \SI{2000}{\kelvin} and \SI{4000}{\kelvin} from \citet{Kitzmann2023}, we searched for H, Fe, Na, Mg, Ca, Li, K, Ti, and V. Although the planet's equilibrium temperature is below \SI{2000}{\kelvin}, we eventually settled for the \SI{4000}{\kelvin} templates to avoid pressure broadening which is relatively stronger in the cooler templates of our species due to higher abundance.

For model comparison, we injected a model of the expected transmission spectrum for \target at the \num{2000} and \SI{3000}{\kelvin} \citep[T$_{\rm eq}=$\SI{1740}{\kelvin};][]{Hellier2019}. The model assumes the planetary atmosphere to be isothermal, in chemical and hydrostatical equilibrium, and at solar metallicity. The planetary parameters are listed in \ref{tab:parameters_W172}. We computed the chemical abundance profiles with \texttt{FastChem} \citep{Stock2018}, which accounts for the variations of molecular weight with altitude, and used these to model the transmission spectrum with \texttt{petitRADTRANS} \citep{Molliere2015}, where we included line absorption from Ca, Cr, Fe, \ch{Fe+}, K, Na, Ti, V. The line lists for the species in the model were taken from \cite{Piskunov1995,ryabchikova_major_2015} \footnote{see \url{https://petitradtrans.readthedocs.io/en/latest/content/available_opacities.html}}, \ch{H2-H2} and \ch{H2-He} were considered for the collision-induced absorption and \ch{H2} and \ch{He} were loaded for Rayleigh scattering. 

To fit for the residual of the planetary obscuration of the stellar disc during transit, we construct an empirical model of Fe by fitting a double Gaussian similar to \citep{Prinoth2022}. The correction steps for the planetary obscuration of the stellar disc are shown in Figure \,\ref{fig:DS_correction}.  
Finally, we apply a high-pass filter to correct for any residual broad-band variations before converting the cross-correlation maps into \kpvsys maps, see \citet{Hoeijmakers2020,Prinoth2022} for extensive discussion.

\subsection{Detections}
\begin{figure}
    \centering
    \includegraphics[width=\linewidth]{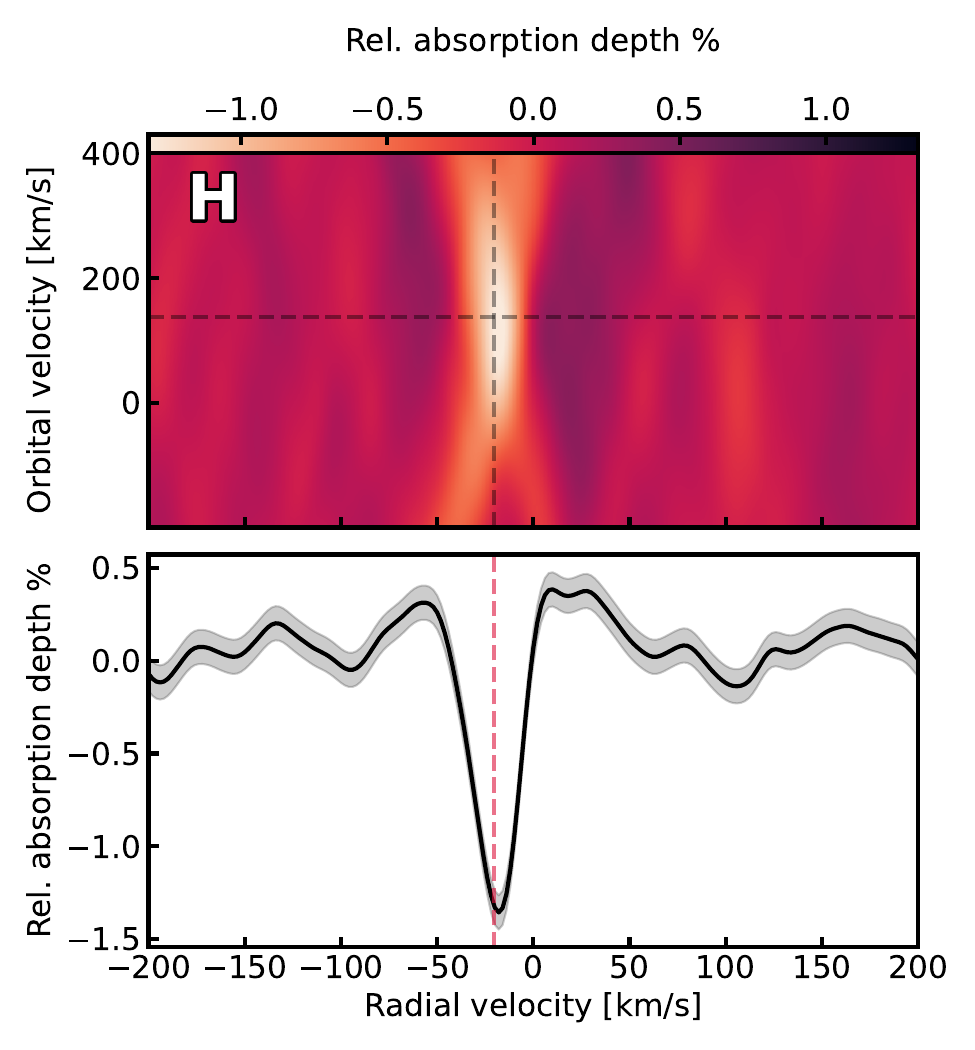}
    \caption{Cross-correlation result for H in the atmosphere of \target. \textit{Top panel}: \kpvsys diagram for H. The dashed grey lines indicate the true orbital and systemic velocities of \vorb\si{\km\per\second} and \vsys\si{\km\per\second}, respectively. The absorption feature appears at the expected location. \textit{Bottom panel:} One-dimensional cross-correlation function at the true orbital velocity. The dashed vertical pink line indicates the true systemic velocity, the same as in the upper panel.}
    \label{fig:kpvsys_H}
\end{figure}

\begin{figure}
    \centering
    \includegraphics[width=\linewidth]{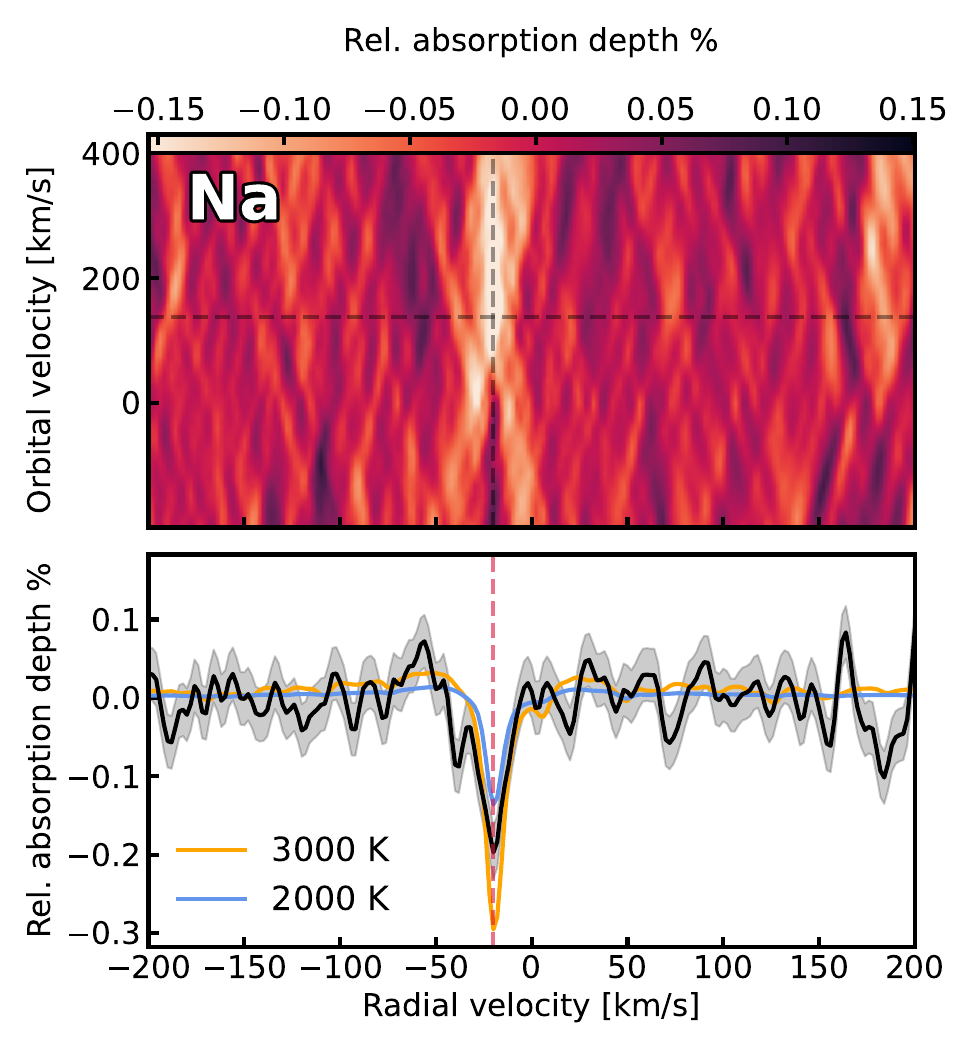}
    \caption{Same as Figure \,\ref{fig:kpvsys_H} but for Na. The blue and orange lines show the predicted relative absorption depth for the models at  \num{2000} and \SI{3000}{\kelvin}, respectively.}
    \label{fig:kpvsys_Na}
\end{figure}

\begin{figure}
    \centering
    \includegraphics[width=\linewidth]{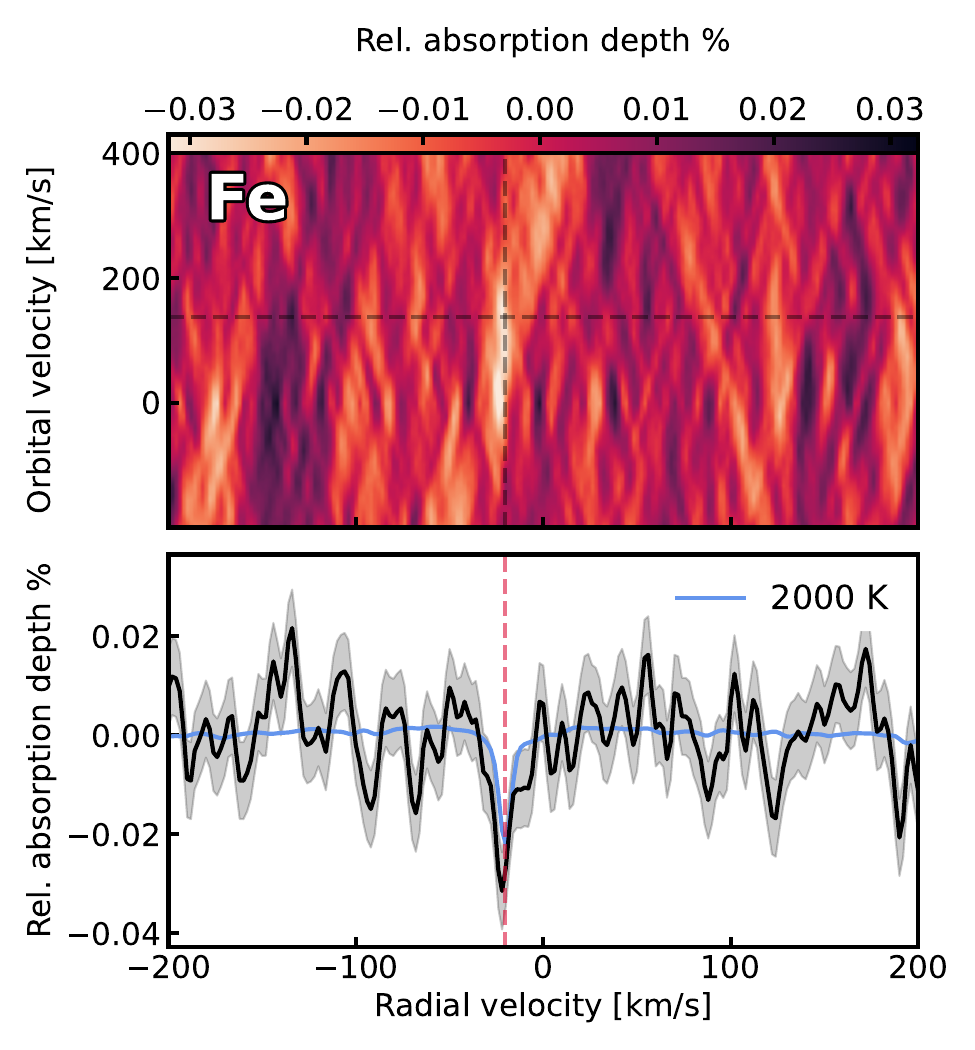}
    \caption{Same as Figure \,\ref{fig:kpvsys_H} but for Fe. The blue shows the predicted relative absorption depth for the model at \SI{2000}{\kelvin}. The model prediction for \SI{3000}{\kelvin} is not shown due to the large predicted absorption depth ($\sim$\num{0.5}\%).}
    \label{fig:kpvsys_Fe}
\end{figure}

Using the cross-correlation technique, we detect H, Na, and Fe in the transmission spectrum of \target. The detections are shown in Figs.\,\ref{fig:kpvsys_H}--\ref{fig:kpvsys_Fe} in \kpvsys space. While the detection of H aligns with the expected orbital and systemic velocity, the detection of Fe shows residuals from removing the signal of the planetary obscuration of the stellar disc during transit. Due to the relatively low orbital velocity of the planet of only \SI{137(2.2)}{\km\per\second}, we, therefore, caution the interpretation of this detection as contamination from the residual of the planetary obscuration during transit may increase the observed signal. However, because the velocity traces of the planet and the obscured stellar disk are in opposing directions, contamination of any signal originating from the planet's atmosphere is minimised. This architecture makes WASP-172~b a favourable system for high-resolution transmission spectroscopy. 

To test whether the signal originates uniformly from in-transit exposures and that it does not appear in out-of-transit exposures, we performed a similar false-positive assessment as described in Section\,\ref{sec:false_pos}. Instead of wavelength space, we use radial velocity space which allows us to perform the false-positive analysis directly on the two-dimensional cross-correlation function. The variation for cross-correlation space is described in detail in Appendix A of \citet{Hoeijmakers2020}. Figures \,\ref{fig:bootstrap_Fe}--\ref{fig:bootstrap_Na} show the result of the false-positive assessment for the detected species.

\begin{figure}
    \centering
    \includegraphics[width=\linewidth]{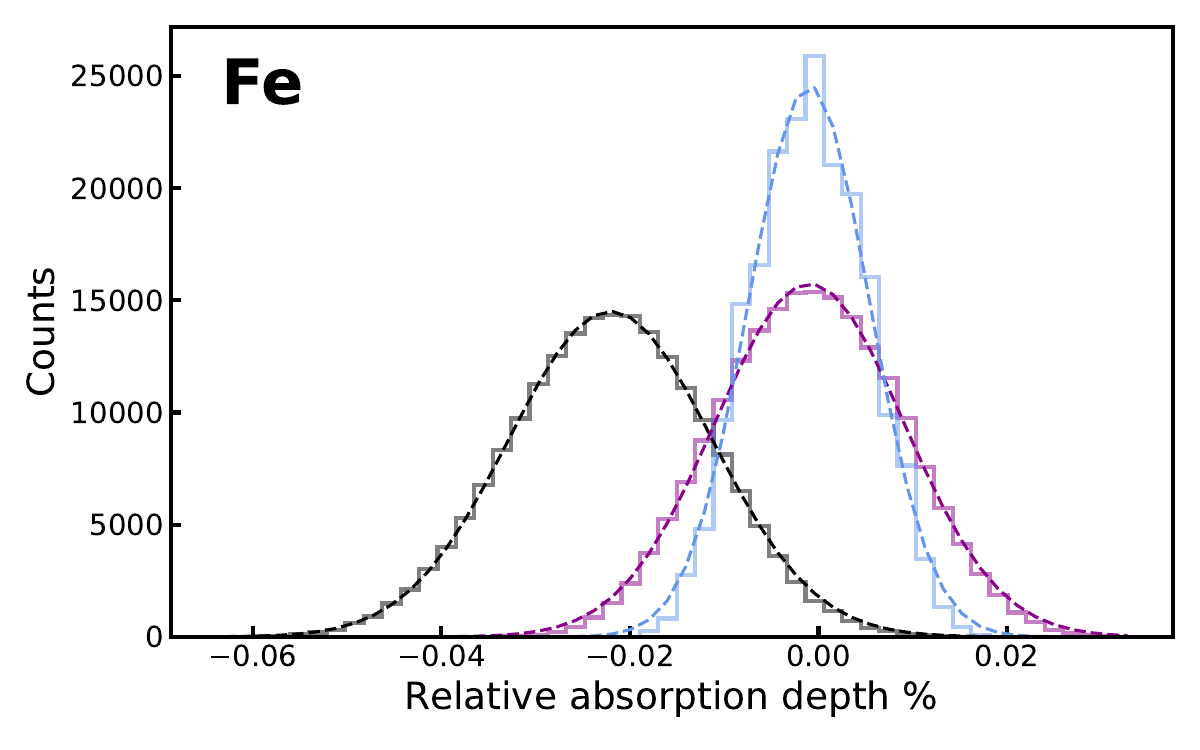}
    \caption{Distribution of bootstrapping analysis for 200,000 random selections. Similar to Figure \,\ref{fig:bootstrap}, the 'in-in' (magenta) and 'out-out' (light-blue) distributions are centred around zero, indicating the absence of the Fe. The 'in-out' distribution in black shows the planetary origin of the feature. For the 'out-out' distribution we only included the seven out-of-transit exposures that were considered in the cross-correlation analysis.}
    \label{fig:bootstrap_Fe}
\end{figure}

\begin{figure}
    \centering
    \includegraphics[width=\linewidth]{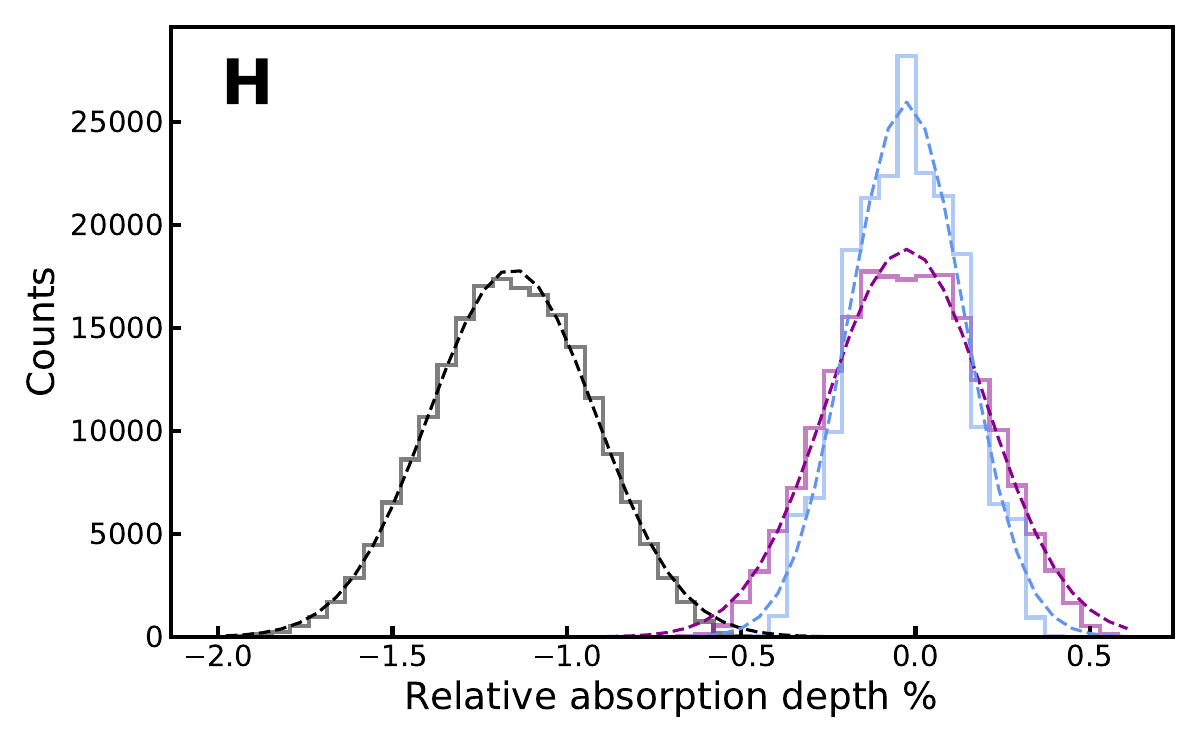}
    \caption{Same as Figure \,\ref{fig:bootstrap_Fe} but for H.}
    \label{fig:bootstrap_H}
\end{figure}

\begin{figure}
    \centering
    \includegraphics[width=\linewidth]{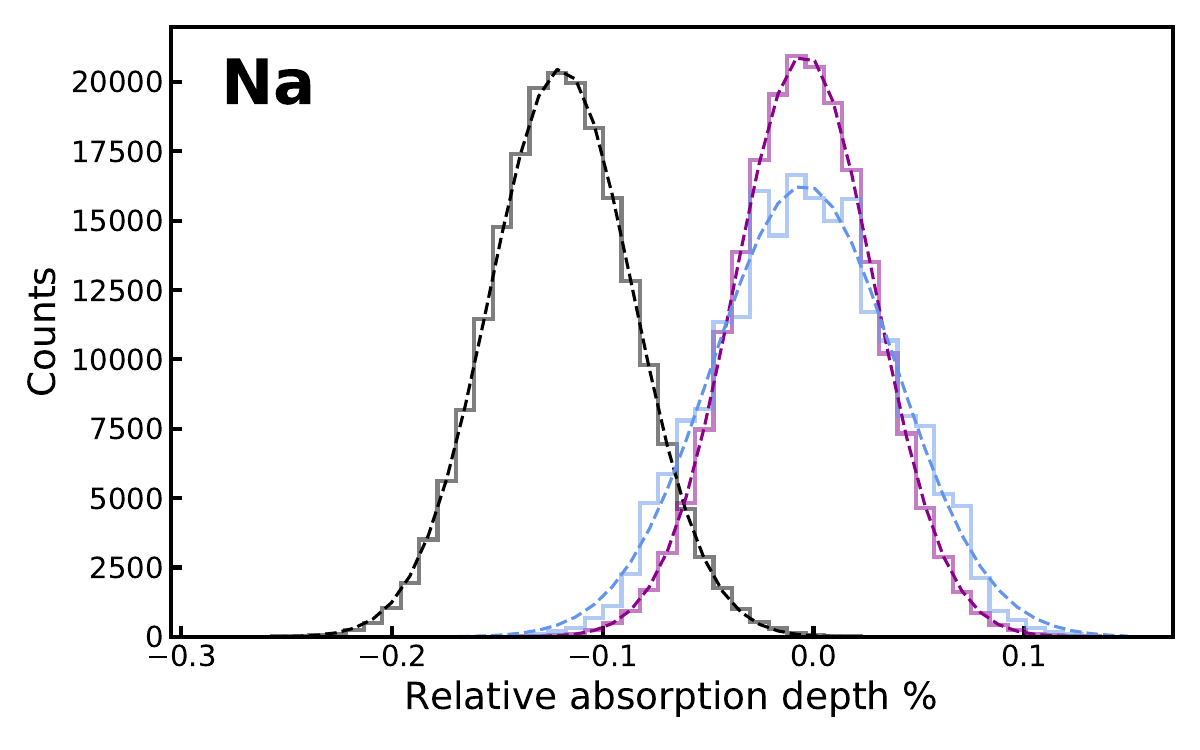}
    \caption{Same as Figure \,\ref{fig:bootstrap_Fe} but for Na.}
    \label{fig:bootstrap_Na}
\end{figure}

\section{Discussion and conclusions}
\label{sec:diss}

\subsection{Atmospheric composition}

\begin{figure}
    \centering
    \includegraphics[width=\linewidth]{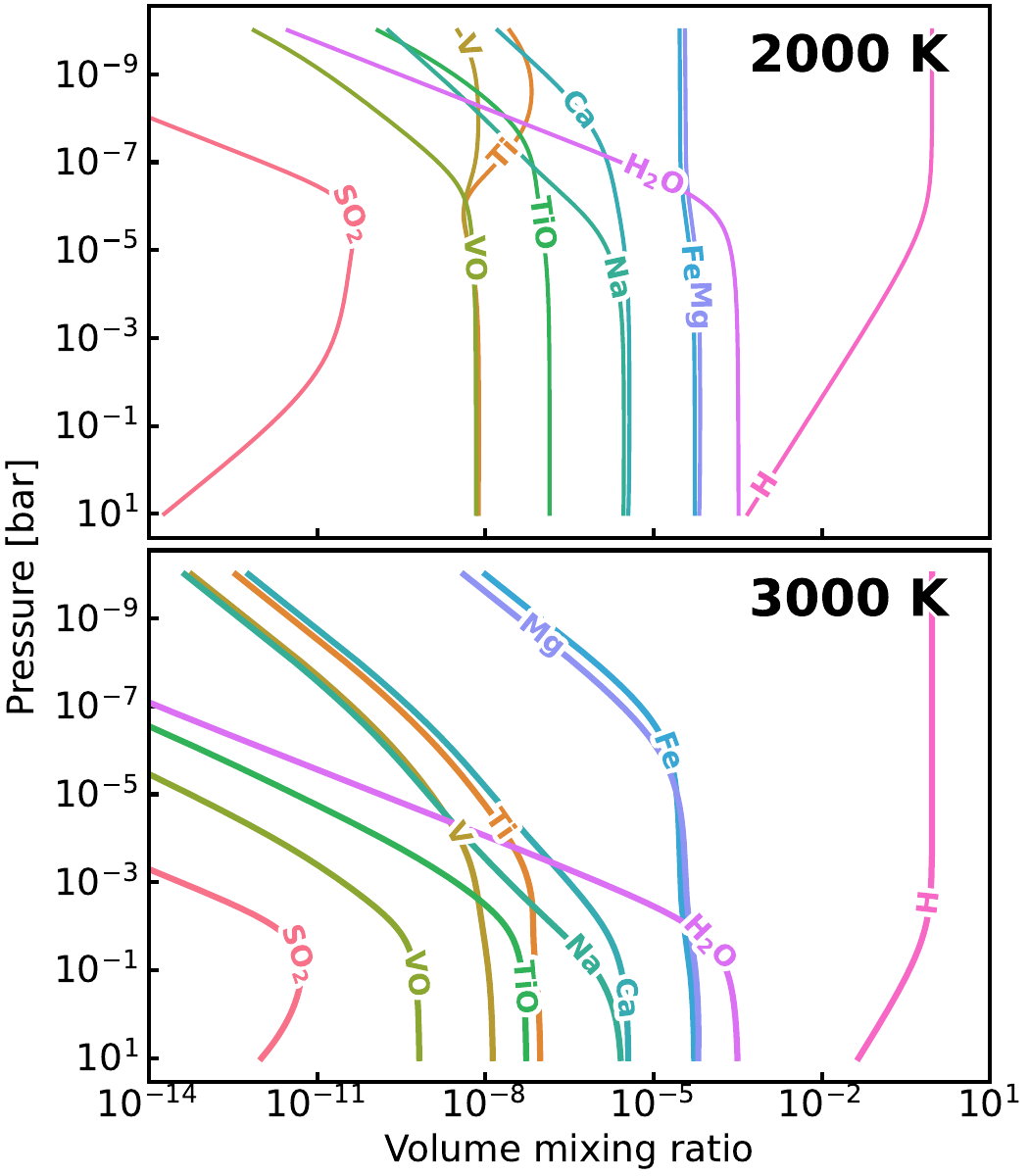}
    \caption{Model of the abundances of key selected species as a function of pressure (inverse altitude) at a temperature of \num{2000} (top) and \SI{3000}{\kelvin} (bottom), assuming thermo-chemical equilibrium and solar metallicity. The abundance profiles were computed with \texttt{FastChem} \citep{Stock2018}.}
    \label{fig:chemistry}
\end{figure}

Fe is known to be a strong absorber in the atmospheres of (ultra-) hot Jupiters and is nowadays routinely detected using the cross-correlation technique \citep[e.g.][]{Hoeijmakers2018,Stangret2020,Hoeijmakers2020,Prinoth2022,kesseli_atomic_2022}. The injected model at a temperature of \SI{2000}{\kelvin} matches the observed absorption depth for Fe, while Na requires a model at a higher temperature. The injected models all have a higher temperature than the equilibrium temperature of this planet (\SI{1740(60)}{\kelvin}) suggesting that the temperature at the terminator exceeds the equilibrium temperature, and/or that the atmosphere is more extended than expected from hydrostatic equilibrium -- frequently observed in hotter ultra-hot Jupiters \citep[e.g.]{Hoeijmakers2019}. Figure \,\ref{fig:chemistry} shows the expected abundances of selected species as a function of pressure (inverse altitude) for both \SI{2000}{\kelvin} and \SI{3000}{\kelvin}. It is expected that Fe and Mg are abundant at relatively higher altitudes, while other metals thermally ionise above approximately 1 $\mu$bar at \SI{2000}{\kelvin} or 1 mbar at \SI{3000}{\kelvin}. Like Na, the absorption spectrum of Mg is dominated by a few strong absorption lines, making the application of cross-correlation less effective than for elements with richer spectra. Fe absorbs over the whole wavelength range of ESPRESSO, but individual lines are intrinsically weaker than the resonant lines of the alkali metals. Even at \SI{2000}{\kelvin}, molecular hydrogen significantly dissociates at pressures below 1 mbar, meaning that the atmosphere at the terminator transitions from molecular to atomic over the altitude range probed by the transmission spectrum. This explains why deep single lines of Na are detectable in the transmission spectrum, while the much richer spectrum of Fe results in a relatively marginal detection. The lines of Fe are intrinsically weaker, meaning that they are formed at lower altitudes where the atmospheric scale height is approximately a factor of two smaller.

Even though the signal of Fe is offset from the system rest-frame velocity (see the velocity-velocity diagram in Figure  \ref{fig:kpvsys_Fe}), due to the relatively small planetary radial velocity (see Figure \,\ref{fig:DS_investiagtion}), the signal of Fe may still be consistent with the rest-frame velocity of the star. This could in theory indicate a stellar origin. To better constrain the velocity, more transit observations would be required to confirm the detection of Fe in the transmission spectrum of \target. If confirmed, the detection of Fe in the transmission spectrum of \target adds to the detections for heavily inflated hot Jupiters, see Figure \,\ref{fig:T_rho}. Importantly, because \target is a relatively cold planet that is not regarded as an ultra-hot Jupiter (typical definitions place the boundary above \SI{2000}{\kelvin}), a robust confirmation of Fe would have implications for our understanding of the chemistry of planets that straddle the hot-ultra-hot Jupiter boundary.

Recently, \cite{Stangret2022} reported the non-detection of Fe in the atmosphere of KELT-17\,b with the HARPS spectrograph, a planet with a higher equilibrium temperature, but less inflated atmosphere. The host star KELT-17\,b is a variable Am-star and subject to pulsations such that a detailed analysis is difficult \citep{Saffe2020}. Fe was similarly not detected in the atmosphere of the hotter but less inflated WASP-19~b \citep{Sedaghati2021} after multiple transit observations with ESPRESSO. This suggests that planets that are less inflated require a higher equilibrium temperature for Fe absorption to be significantly detectable. The planet with the lowest temperature and reported Fe detection is HD~142096~b \citep{Ishizuka2021}, an inflated hot Jupiter, like \target, with a dayside temperature of $\sim 2100~K$. However, this detection is only confirmed at the $2.8 \sigma$ level. From both similar targets in the literature and this work it is, therefore, unclear whether neutral iron is present in inflated, but cooler, hot Jupiters, and follow-up observations are crucial.

Na and H are detected with high significance both using narrow-band transmission spectroscopy and the cross-correlation technique.

\subsection{Atmospheric dynamics}
\label{sec:dynamics}

\begin{figure*}[htb]
\resizebox{\textwidth}{!}{\includegraphics[trim=0.5cm 0.0cm 0.5cm 0.0cm]{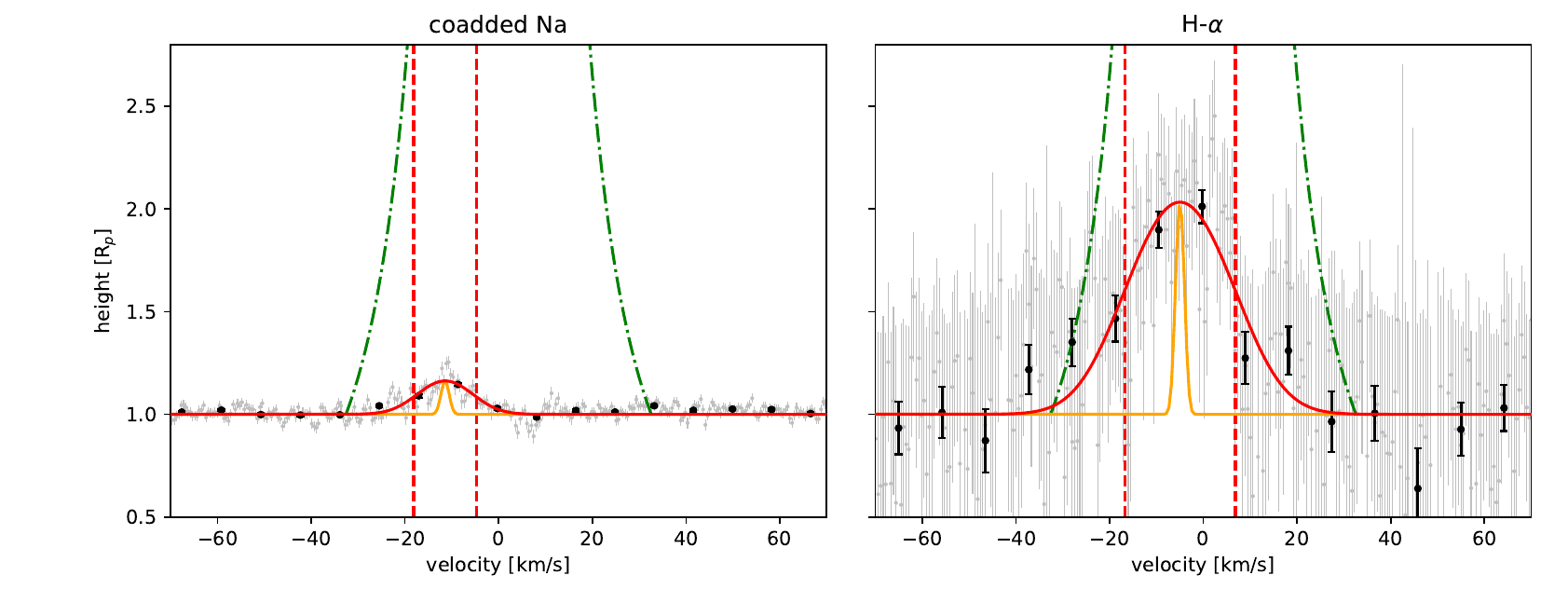}}
        \caption{The coadded sodium doublet and the H-$\alpha$ line in velocity space as a function of altitude above the white light radius. The Gaussian fits together with the FWHM (dashed, vertical) is shown in red, the ESPRESSO line spread function in yellow, and the escape velocity as a function of height in dashed-dotted green.}
        \label{fig:vel_plot}
\end{figure*}

In the two species detected via narrow-band transmission spectroscopy a clear blue shift is detectable. The sodium doublet in air is expected at $5889.950$ and $5895.924~\AA$, but the Gaussian fit in our data sets the centre of the lines at $5889.728\pm0.011$ and $5895.663\pm 0.016~\AA$, respectively. This corresponds to a blueshift towards the observer of $11.3$ and $13.2~\kms$, respectively. Both sodium lines should be shifted in the same way and the difference in observed blueshift highlights the ambiguity of the line shape for the study of atmospheric studies. A dedicated observation of this target geared towards atmospheric dynamics will remedy this shortcoming. Sodium traces the atmosphere from the lower atmosphere at the micro-bar pressure level up to the start of potential inversion layers in a thermosphere\citep{Wyttenbach2015,Seidel2020}, while H-$\alpha$ as the main hydrogen line traces the atmosphere in its full vertical extension. H-$\alpha$ is expected in air at $6562.810~\AA$ but is detected here at $6562.705\pm0.01~\AA$. This shift corresponds to a movement towards the observer at $4.8~\kms$. 
\noindent Assuming that WASP-172~b is tidally locked, the rotational velocity of the atmosphere is approximately $1.5~\kms$. In broad terms, integrated over the entire atmosphere as traced by H-$\alpha$ and taking into account the planetary rotation, the atmosphere shows a movement from the hot dayside facing the star towards the cooler nightside of $\sim 3-6~\kms$. The sodium doublet, which traces largely lower layers of the atmosphere in comparison, shows the same overall movement at higher wind speeds between $\sim 10-14~\kms$. The difference in wind speeds at the different altitude levels hints at high-velocity day-to-night side winds which might be localised in jet streams at the microbar level and above, as seen for a wide range of Jupiter-like planets, from emission features on the hottest ever found planet, KELT-9~b \citep{Pino2022} to ultra-hot Jupiters \citep{Ehrenreich2020,Kesseli2021,Seidel2021,Gandhi2022,Brogi2023,Seidel2023,Gandhi2023} to the cooler end of hot Jupiters, for example HD~189733~b \citep{Seidel2020}. 

Compared to the line spread function of the instrument, both the sodium doublet and the H-$\alpha$ line are broadened significantly with a FWHM of $13.4~\kms$ and $23.6~\kms$, respectively (see Figure \ref{fig:vel_plot}). This broadening is indicative of either a super-rotational wind in the entire atmosphere or a radial, vertical wind pattern moving outwards. However, despite the low mass of the planet, neither species is moving beyond the escape velocity, indicating a stable atmosphere. However, understanding atmospheric dynamics from Doppler shifts requires a high level of confidence not only in the detection but also in the line shape and position. As stated before for the two lines of the sodium doublet, the uncertainty on the shift is most likely at the order of various $\kms$, and without further data, it is ambiguous if the observed resolved lines truly are at significantly different velocities. It is encouraging that similar shifts are seen with the cross-correlation technique, which implies that there was no artificial shift introduced by the data reduction method of the narrow-band transmission spectra. Nonetheless, follow-up observations with a more reliable out-of-transit baseline to prove the repeatability of the observations are crucial for a more in-depth study of the atmospheric dynamics of WASP-172~b.

 \subsection{Future avenues}

\begin{figure}
   \resizebox{\linewidth}{!}{\includegraphics[trim=3cm 1.0cm 1cm 0.0cm]{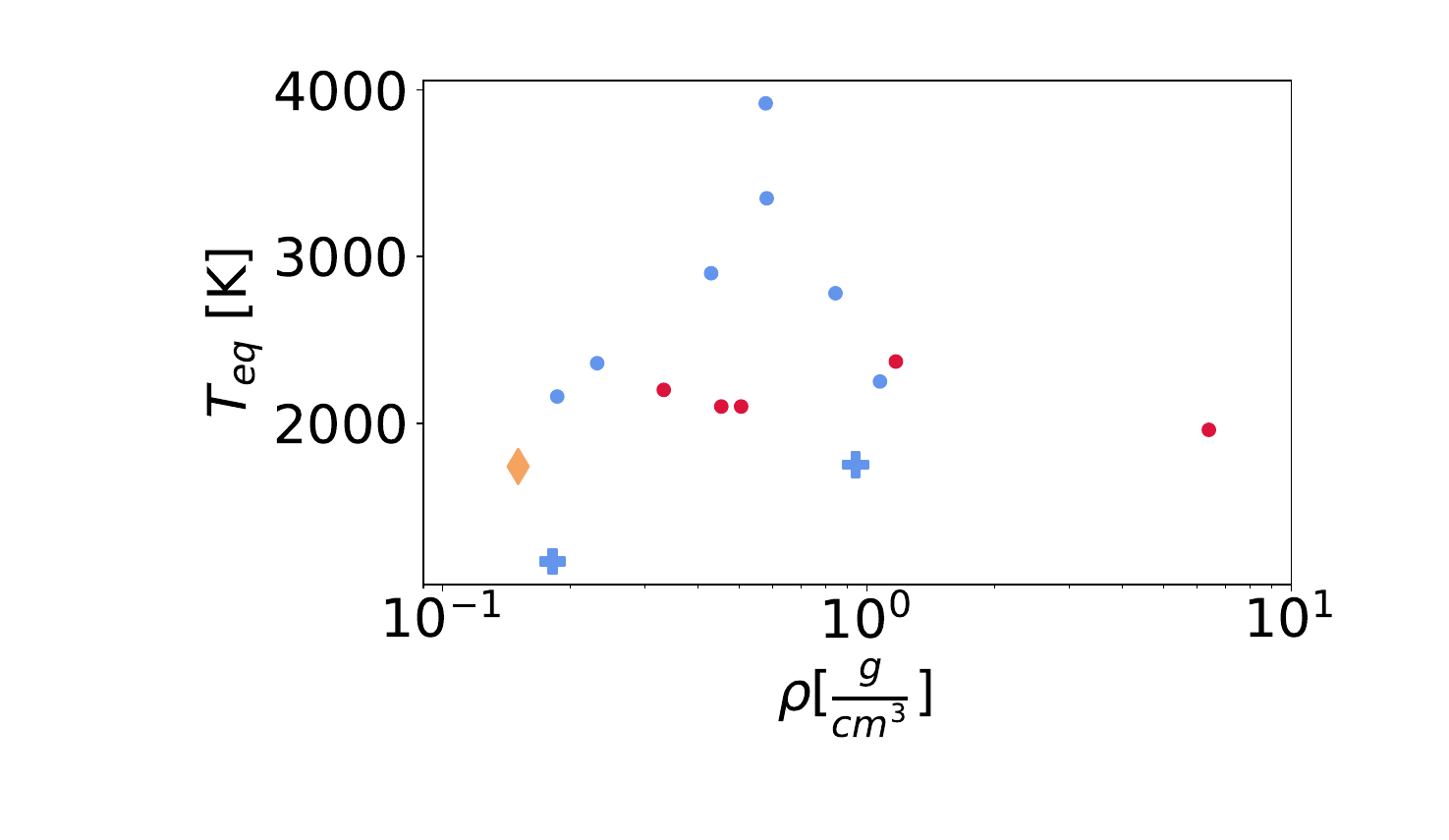}}
    \caption{WASP-172~b in temperature-density space together with all currently known Fe detections in blue for exoplanets with confirmed Fe, as well as known upper limits in red. We have limited the data to exoplanets with a well-constrained mass. Already observed JWST targets in the lower temperature range are marked with crosses (HD~149026~b, WASP-39~b), and WASP-172~b is marked as a brown diamond. From \url{http://research.iac.es/proyecto/exoatmospheres/table.php}, retrieved August 2023.}
    \label{fig:T_rho}
\end{figure}

\noindent Considering that we only have one transit available and cannot confidently resolve the line shape, it is not possible to properly understand which kind of wind pattern generates the offsets we see here and a day-to-night side wind is as possible as a one-sided jet or a super-rotational wind stream. However, the clear offsets in multiple lines, as well as the clear detection in CCF make WASP-172~b a prime candidate for time-resolved transit observations to bridge the transition from hot to ultra-hot Jupiters. In Figure \ref{fig:T_rho}, WASP-172~b is shown together with all currently known exoplanets with confirmed or tentative iron detections that also have well-constrained masses and, therefore, densities. The two other exoplanets in the same lower temperature range of hot Jupiters, HD~149026~b, and WASP-39~b, are marked with crosses and have been, coincidentally, observed with JWST.

 WASP-39~b has provided an outstanding target with a clear detection of \ch{H2O} \citep{Alderson2023} as well as the first detection of photochemical processes in an exoplanet via \ch{SO2} \citep{Alderson2023,Tsai2023}. HD~149026~b has similarly shown high metal enrichment \citep{Bean2023}, highlighting the potential of the similar WASP-172~b for follow-up with ground and space-based facilities to broaden our understanding of the transition from ultra-hot to cooler hot Jupiters.

%----------------------------------------------------------------------------------------
%       ACKNOWLEDGEMENTS
%----------------------------------------------------------------------------------------
\begin{acknowledgements}
The authors acknowledge the ESPRESSO project team for its effort and dedication in building the ESPRESSO instrument. This work relied on observations collected at the European Organisation for Astronomical Research in the Southern Hemisphere. S.A.\ acknowledges the support from the Danish Council for Independent Research through a DFF Research Project 1 grant, No.\ 2032-00230B.

\end{acknowledgements}

%----------------------------------------------------------------------------------------
%       REFERENCE LIST
%----------------------------------------------------------------------------------------
%
\bibliographystyle{aa} % style aa.bst
\bibliography{WASP172}

%----------------------------------------------------------------------------------------
%       APPENDIX
%----------------------------------------------------------------------------------------
%
\appendix
\onecolumn
\section{Cross-correlation map}

\begin{figure*}[htb]
    \centering
    \includegraphics[width=\linewidth]{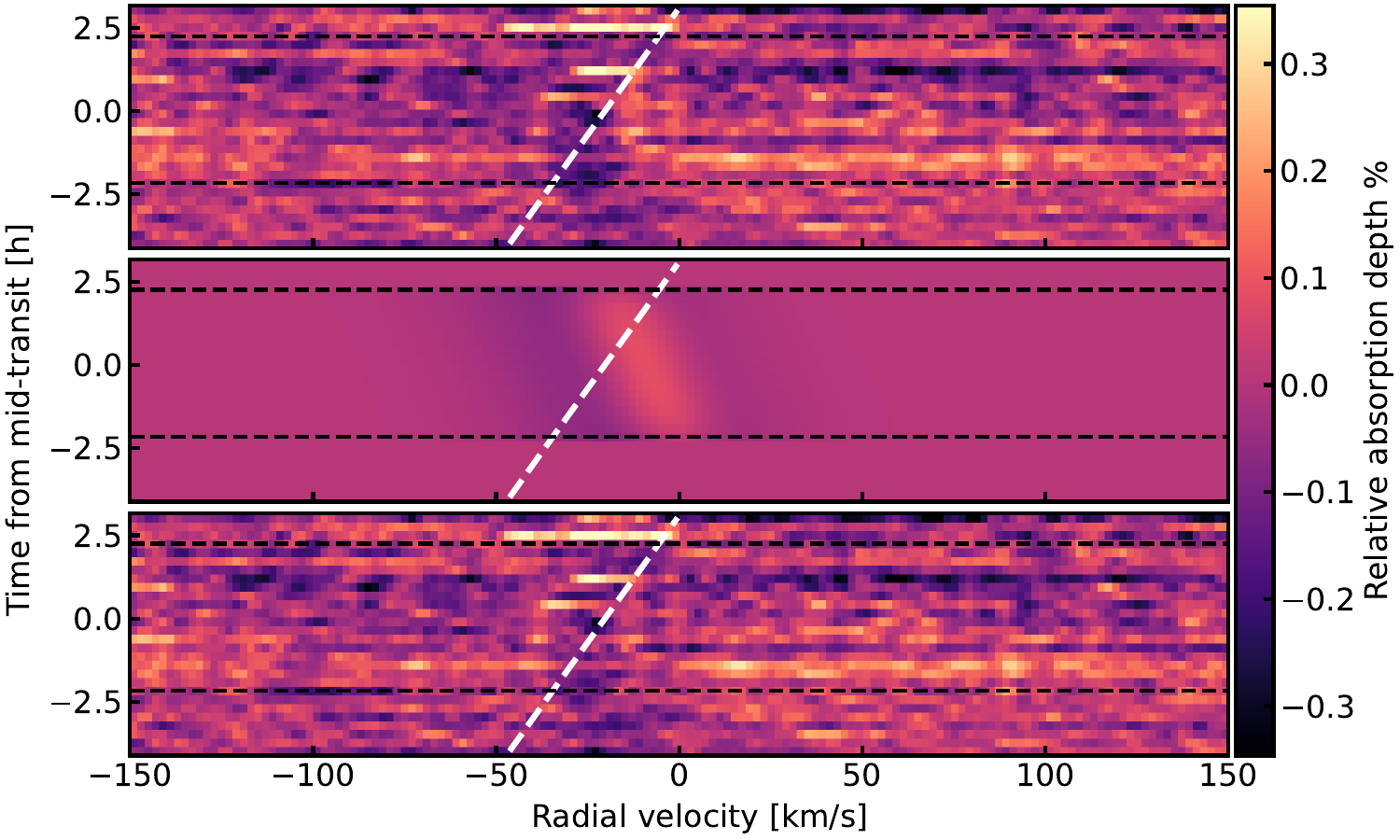}
    \caption{Correction of the residual of the planetary obscuration of the stellar disc (Doppler shadow). The planetary trace is shown in white. The black dashed lines indicate the start and the end of the transit. \textit{Top panel:} Two-dimensional cross-correlation function. \textit{Middle panel:} Model of the residual of the planetary obscuration of the stellar disc with two Gaussian components. \textit{Bottom panel:} Two-dimensional cross-correlation function after correction.}
    \label{fig:DS_correction}
\end{figure*}

\end{document}